\documentclass[aps,pre,twocolumn,footinbib,superscriptaddress]{revtex4-2}  
\usepackage{graphicx}  
\usepackage{dcolumn}   
\usepackage{bm}        
\usepackage{amssymb}   
\usepackage{amsmath}   
\usepackage{extarrows}

\newcommand{\Ho}{\hat{H}}
\newcommand{\Uo}{\hat{U}}
\newcommand{\no}{\hat{n}}
\renewcommand{\ao}{\hat{a}}
\renewcommand{\aa}{\hat{a}^\dag}

\begin{document}

\title{Design and characterization of a quantum heat pump in a driven 
quantum gas}
\author{Arko~Roy} 
\email[Electronic address: ]{arko.roy@unitn.it}
\affiliation{Max-Planck-Institut f{\"u}r Physik komplexer Systeme, 
N{\"o}thnitzer Stra\ss e 38, 01187 Dresden, Germany}
\affiliation{INO-CNR BEC Center and Dipartimento di Fisica, Universit\`a di Trento, 38123 Trento, Italy}
\author{Andr{\'e}~Eckardt} 
\email[Electronic address: ]{eckardt@tu-berlin.de}
\affiliation{Max-Planck-Institut f{\"u}r Physik komplexer Systeme, 
N{\"o}thnitzer Stra\ss e 38, 01187 Dresden, Germany}
\affiliation{Technische Universität Berlin, Institut f\"ur Theoretische Physik, Hardenbergstra{\ss}e 36, 10623 Berlin, Germany}

\date{\today}
\begin{abstract}
We propose the implementation of a quantum heat pump with ultracold atoms. It is based on two periodically driven coherently coupled quantum dots using ultracold atoms. Each dot possesses two relevant quantum states and is coupled to a fermionic reservoir. The working principle is based on energy-selective driving-induced resonant tunneling processes, where a particle that tunnels from one dot to the other either absorbs or emits the energy quantum $\hbar\omega$ associated with the driving frequency, depending on its energy. We characterize the device using Floquet theory and compare simple analytical estimates to numerical simulations based on the Floquet-Born-Markov formalism. In particular, we show that driving-induced heating is directly linked to the micromotion of the Floquet states of the system.

\end{abstract}

\maketitle

\section{Introduction}
The miniaturization of heat engines and pumps to systems consisting of a few 
relevant quantum states only \cite{kosloff_14,benenti_17} and their description 
in terms of quantum thermodynamics \cite{campisi_11,vinjanampathy_16,brandao_15,
gemmer2004} constitutes a fascinating and active field of research. In this 
context the implementation and investigation of such devices with 
ultracold neutral atoms in tailored light-shift potentials defines a promising
direction of research. Especially the recently developed quantum-gas 
microscopes, where digital mirror devices are employed for microstructuring 
almost arbitrary potential landscapes with high resolution both in space and 
time \cite{ott_16,kuhr_16,cheuk_15,alberti_16,mcdonald_19,mazurenko_19}, 
provide an interesting platform for this goal. One advantage of atomic quantum
gases in optical potentials is that they provide extremely clean conditions for
studying the fundamental properties of quantum engines and pumps, since they do
not suffer from dissipation induced by the coupling to phonons or due to 
radiative loss, as it is typically present in electronic systems. First 
experiments in this direction include the creation of a heat engine
\cite{brantut_13} as well as local probes for thermometry in ultracold gases
\cite{mayer_19,bouton_20,hohmann_16}. 
Moreover, the implementation of a heat pump could be also of practical 
use for reaching lower temperatures.

\begin{figure}
  \includegraphics[width=0.7\linewidth]{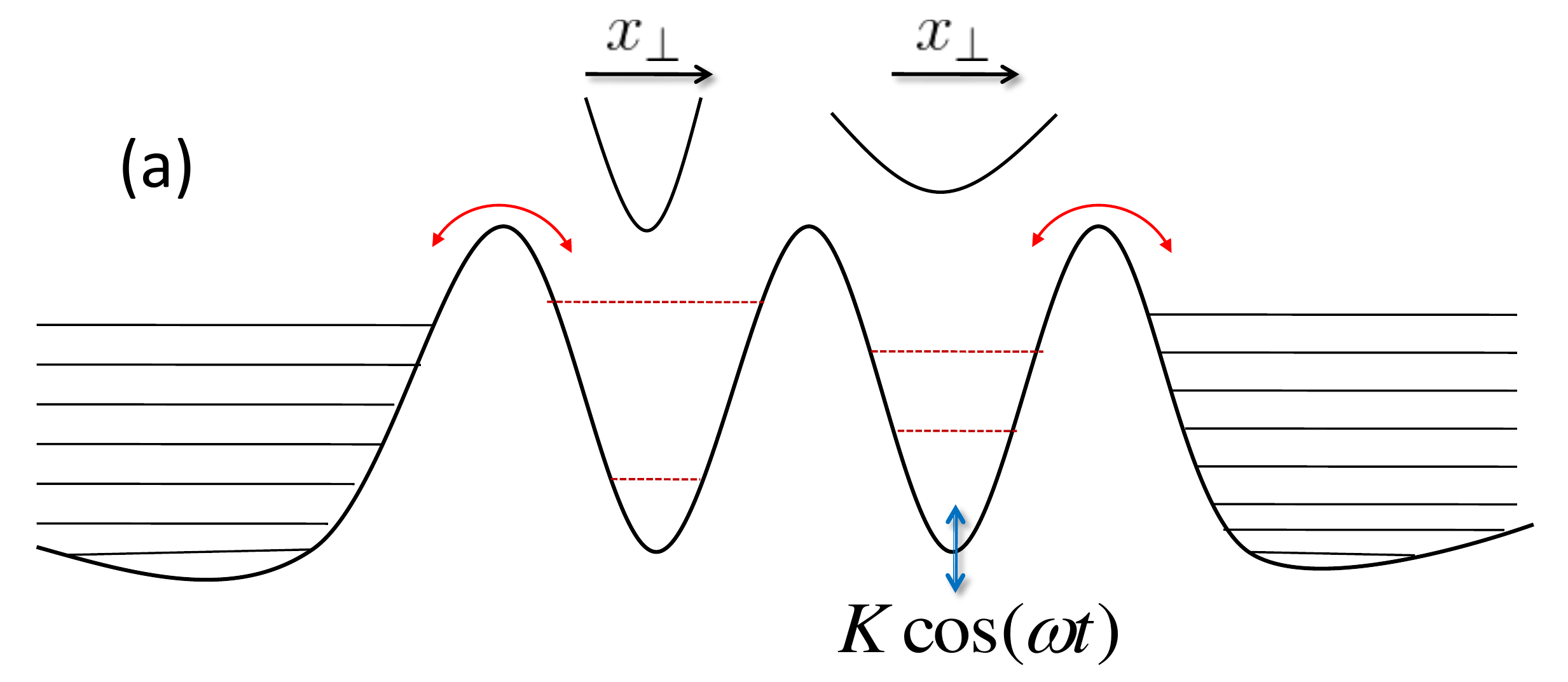}
  \includegraphics[width=0.8\linewidth]{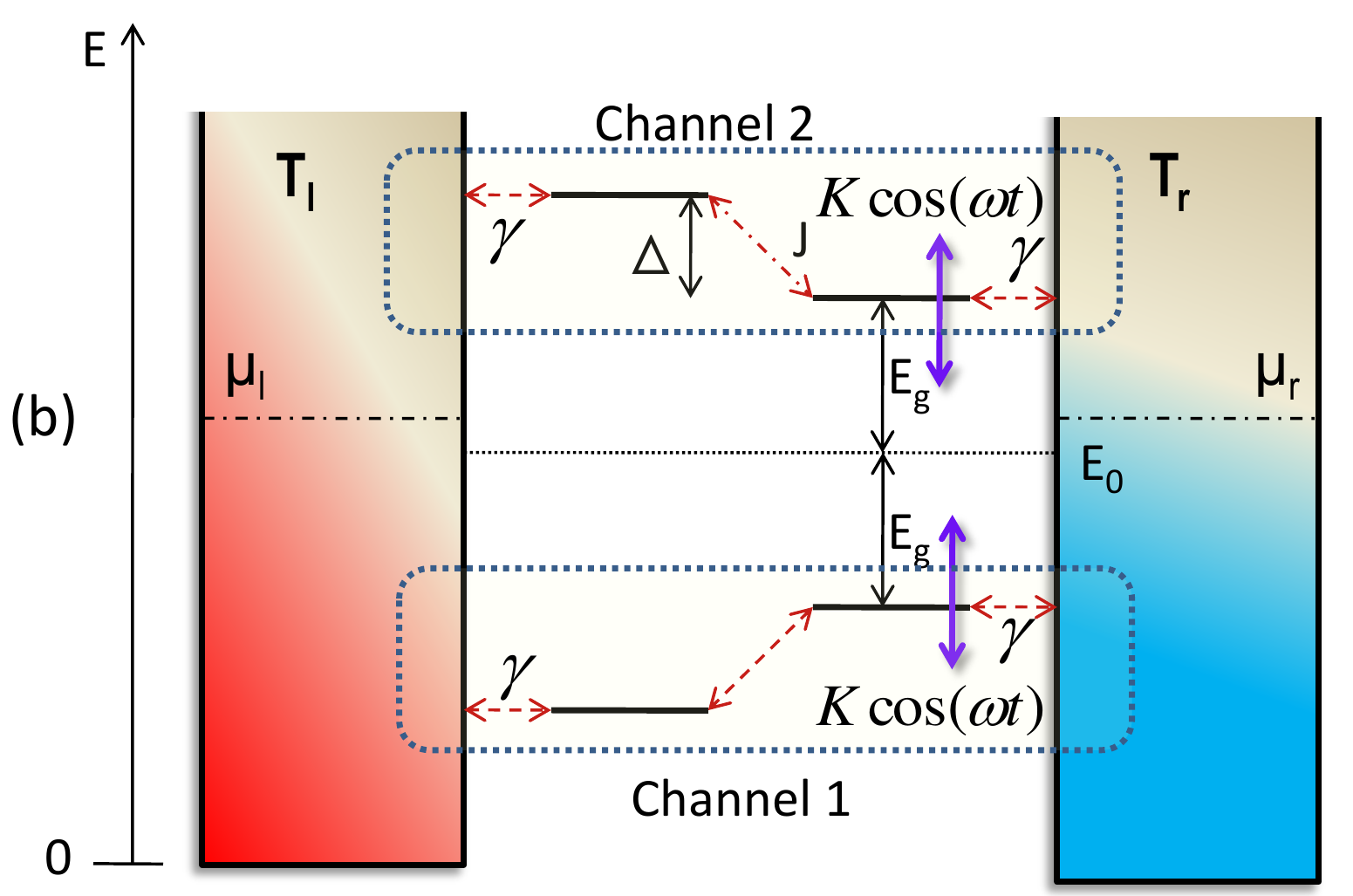}
  \caption{(a) Sketch of the potential landscape used to build the heat pump. 
	(b) Open-system model used to describe the device.}
  \label{potential}
\end{figure}

In this paper, we design, characterize and propose to implement a quantum heat
pump in an optically microstructured quantum gas. The starting point is a setup
as it is realized by the Zurich group, where two reservoirs are coupled by a 
structured channel \cite{brantut_13,krinner_14,krinner_17,hausler_17}. The 
device itself is based on the potential landscape sketched in 
Fig.~\ref{potential}(a). It consists of two coupled quantum dots, to be labeled 
$l$ (left) and $r$ (right), each coupled to a larger fermionic system 
and each hosting two relevant single-particle levels, to be labeled by 1 (lower)
and 2 (upper) [Fig.~\ref{potential}(a)]. The upper levels shall correspond to 
excitations transverse to the sketched potential. This brings various 
advantages: the energies of the individual levels can easily be tuned by the 
local transverse potential, the tunnel coupling between the upper levels is 
comparable to that of the lower ones, and unwanted tunneling between the upper 
and lower level of different dots is forbidden by opposite transverse parity. 
Finally, we assume a time-periodic energy modulation $K\cos(\omega t)$ for 
particles in the right dot with amplitude $K$ and frequency $\omega$.
The proposed setup can be realized using available experimental techniques: Two 
fermionic reservoirs coupled by tailor-made potential landscapes have already 
been realized by the Zurich group \cite{brantut_13,krinner_14,krinner_17,
hausler_17}. Furthermore, the combination of digital mirror devices with 
high-resolution optics has become a standard technique for engineering 
light-shift potentials with high spatio-temporal resolution \cite{kuhr_16}.

A driven double-quantum-dot structure similar to the one considered here was 
proposed also for electrons in a driven heterostructure \cite{ReyEtAl2007}.
It was modeled as a one-dimensional system with stepwise constant potentials, 
which allowed for a treatment using a Floquet transfer-matrix approach
\cite{Wagner1995} and which differs from the system considered here by the 
fact that it does not involve transverse degrees of freedom. Different from 
this previous work, our analysis presented below is based on 
Floquet-Born-Markov theory \cite{bluemel_91,kohler_97,hone_09,vorberg_15}. 
Another difference consists in the fact that we are assuming reservoirs of finite 
size, as they are relevant for a quantum-gas system, and study their time 
evolution. The latter reveals the interplay between cooling via energy pumping 
on the one hand, which for (suitable parameters) dominates on short times, 
and heating that we relate to driving induced micromotion on the other, which 
eventually will dominate in the long-time limit.  
The fact that the working 
mechanism of our system will be based on the transport of particles (fermions)
between the dots and the reservoirs, distinguishes it, moreover, from very 
recently proposed periodically driven heat pumps based on energy exchange 
\cite{Charalambous_2019,RieraCampeny_2019} and from an incoherently driven 
double-dot device~\cite{cleuren_12}.

\section{Basic idea}
The model that we employ to describe the system is sketched in 
Fig.~\ref{potential}(b). We treat the fermionic systems coupled to the left and 
right quantum dot as thermal reservoirs characterized by temperatures $T_l$, 
$T_r$ and chemical potentials $\mu_l$, $\mu_r$. For simplicity, we assume (i)
the reservoirs (which we label by the dot they are coupled to, $d=l,r$) to have 
a density of states $D_d(E)=\rho_d\theta(E)$ taking a constant value $\rho_d$  
above their minimum energy $E=0$ ($\theta$ denotes the step function) and (ii) 
the particle exchange between reservoirs and dots via tunneling to be captured 
by an energy independent parameter~$\gamma$. 
However, this choice is 
not essential.

Within the double-dot structure, for zero driving ($K=0$) the lower and the upper
state shall be arranged symmetrically with respect to the energy $E_0$ in both 
wells and they shall be separated by $2(E_g+\Delta)$ and $2E_g$ in the left and
the right well, respectively, with $\Delta,E_g>0$. 
Moreover, particles can tunnel ``horizontally'' between 
the lower and the upper pair of levels with matrix element $-J$, whereas
``diagonal'' tunneling between the lower level of one dot to the upper one of the
other dot is suppressed by the opposite parity of the transverse wave-function. 
Thus, the lower and the upper pair of levels form two individual channels, 1 and
2, individually connecting both reservoirs. All in all, the double-dot system is 
described by the Hamiltonian
\begin{eqnarray}
\Ho(t) &=&  (E_0+E_g+\Delta)\no_{2l} +  [E_0+E_g+K\cos(\omega t)] \no_{2r}
\nonumber\\	&&
					+\, (E_0-E_g-\Delta)\no_{1l}
					  + [E_0 - E_g+K\cos(\omega t)] \no_{1r} 
\nonumber\\&&
					-\,J (\aa_{1r}\ao_{1l} + \aa_{2r}\ao_{2l} + \text{h.c.}), 
\end{eqnarray}
where h.c.\ stands for hermitian conjugate and $\ao_{cd}$ and $\no_{cd}$, with dot 
label $d=l,r$ and channel label $c=1,2$, denote the annihilation and number 
operators for spinless fermions in the four levels.  

In the following, we will assume $J\ll\Delta$, so that tunneling between the 
dots is energetically suppressed in the undriven system. However, we will also 
assume that the driving frequency is tuned to resonance 
with this offset, $\Delta = \nu\hbar\omega$ with integer $\nu$, so that 
coherent tunneling can be induced as a ``photon''-assisted process. In this 
way, one can achieve a situation, where fermions of energy $E_0+E_g$ absorb 
$\nu$ energy quanta $\hbar\omega$ from the drive, when passing from the right 
to the left reservoir via channel 2, whereas fermions at the lower energy
$E_0-E_g$ emit the energy $\nu\hbar\omega$ into the drive, when moving from 
right to left via channel 1. Together with the reverse left-to-right 
processes, one can immediately see that a steady-state situation without net 
particle and energy flow between both reservoirs can be given by a 
configuration, where $\mu_l=\mu_r=E_0$ and 
\begin{equation}\label{eq:Tratio}
\frac{T_l}{T_r}= \frac{E_g+\Delta}{E_g} \equiv a > 1.
\end{equation}
Namely, in this case the Fermi-Dirac distribution 
$f_{T,\mu}(E)=\{\exp[(E-\mu)/T]+1\}^{-1}$ of the right reservoir at energies
$E_0\pm E_g$ equals that of the left reservoir at energies $E_0\pm(E_g+\Delta)$,
respectively. Therefore, one can expect that the driven double-dot acts as a 
quantum heat pump, transferring energy from the colder right to the hotter left 
reservoir, as long as
\begin{equation}\label{eq:pumpregime}
\frac{T_l}{a} \lesssim T_r < T_l .
\end{equation}
Note that Eqs.~(\ref{eq:Tratio}) and (\ref{eq:pumpregime}) are simple 
estimates, only. Deviations from them will arise due to ``photon''-assisted 
tunneling processes between the levels $E_0\pm E_g$ of the driven right quantum 
dot and states of the right reservoir at energies $E_0\pm E_g+m\hbar\omega$, 
during which integer numbers $m\ne0$ of energy quanta $\hbar\omega$ are 
absorbed or emitted by the drive. Furthermore, also the level splitting within 
each channel, as it is induced by resonant tunneling, alters this simple picture.

\section{Open-system approach}
Each channel corresponds to a driven two-level system. In the ``rotating'' 
reference frame, obtained by integrating out the potential offsets
$\hat{V}=\sum_c s_c\nu\hbar\omega\no_{cl}+K\cos(\omega t)\no_{cr}$ between both 
wells of each channel, with sign $s_c\equiv(-1)^c$, 
the tunneling parameter becomes time dependent. Namely, after a gauge 
transformation $\Ho'=\Uo^\dag\Ho\Uo-i\hbar\Uo^\dag \frac{d}{d t}\Uo$ with the
time-periodic unitary operator 
$\Uo(t)=\exp\big[-i\int_0^t\!dt\,\hat{V}(t)/\hbar\big]$, we arrive at 
\begin{eqnarray}
\Ho'(t) &=& \sum_{c=1,2} \Big[(E_0+s_cE_g)\no_c 
\nonumber\\&&
					-\,J \Big(e^{i(\alpha\sin(\omega t)-s_c\nu\hbar\omega t)}
					\aa_{cr}\ao_{cl}	+\text{h.c.}\Big) \Big],
\end{eqnarray}
where we have introduced the total channel occupations $\no_c=\no_{cl}+\no_{cr}$
as well as the dimensionless driving amplitude $\alpha\equiv K/(\hbar\omega)$. 
In the high-frequency limit $J\ll\hbar\omega$, we can average this rapidly varying 
phase factor over one driving period to obtain the effective time-independent
Hamiltonian in rotating-wave approximation,
\begin{equation}
\Ho_\text{eff} = \sum_{c=1,2} \Big[(E_0+s_cE_g)\no_{c}
					-J_c^\text{eff}(\aa_{cr}\ao_{cl}	+\text{h.c.}) \Big].
\end{equation}
Here $J_c^\text{eff}=J\mathcal{J}_{s_c\nu}(\alpha)$ denotes the effective tunneling 
matrix element, where $\mathcal{J}_k(x)$ denotes the
$k$th-order Bessel function of the first kind (see, e.g., Ref.~\cite{eckardt_17} 
for details). Diagonalizing $\Ho_\text{eff}$ Hamiltonian and transforming the 
eigenstates back to the original frame of reference, one obtains the time-periodic 
single-particle Floquet modes
$|u_{c\pm}(t)\rangle = \big(e^{-is_c\nu\omega t}|cl\rangle 
\pm e^{-i\alpha\sin(\omega t)}|cr\rangle \big)/\sqrt{2}$, 
with quasienergies $\varepsilon_{c \pm}=E_0+s_c E_g\pm J_c^\text{eff}$, where
$|cd\rangle=\aa_{cd}|\text{vac}\rangle$ with vacuum $|\text{vac}\rangle$. For 
non-interacting fermions, we also define time-periodic Floquet-Fock states
$|\{n_{c\pm}\}(t)\rangle$ characterized by sharp occupation numbers
$n_{c\pm}$ of the single-particle Floquet modes.

Let us treat the double-quantum dot as an open system coupled to the left and 
the right reservoir [Fig.~\ref{potential}(b)]. In the limit where the coupling 
to the reservoirs, $\gamma$, becomes small compared to the quasienergy level 
splitting $\sim |J_c^\text{eff}|$), the system approaches a quasi-steady state 
described by a time-periodic density matrix $\hat{\rho}(t)=\sum_{\{n_{c\pm}\}}
p_{\{n_{c\pm}\}}|\{n_{c\pm}\}(t)\rangle\langle\{n_{c\pm}\}(t)|$, which is 
diagonal with respect to the time-periodic Floquet-Fock states \cite{bluemel_91,
kohler_97, hone_09,vorberg_15}. The diagonal elements are given by 
time-independent probabilities $p_{\{n_{c\pm}\}}$, which are determined by the 
rates $(1-n_{c\pm })R^{*}_{c\pm}$ and $n_{c\pm} R^{\dag}_{c\pm}$ for the gain
(``birth'' $*$) and the loss (``death'' $\dag$) of a fermion in state
$c\pm$, respectively. The rates have contributions from both reservoirs 
($d=l,r$), $R^{\eta}_{c\pm}=R^{\eta l}_{c\pm}+R^{\eta r}_{c\pm}$ with $\eta=*,\dag$, and can be obtained using Floquet-Born-Markov theory in
combination with the secular approximation 
\cite{bluemel_91,kohler_97,hone_09,vorberg_15}. They are given by a sum of 
golden-rule type terms describing processes where the system exchanges $m$ 
energy quanta $\hbar\omega$ with the drive, 
$R_{c\pm}^{\eta d}=\sum_mR_{c\pm}^{\eta d(m)}$ with 
\begin{equation}\label{eq:Rm}
R_{c\pm}^{\eta d(m)}=\frac{2\pi}{\hbar} |\gamma_{cd,\pm}^{(m)}|^2 
D_d(\varepsilon_{c\pm}+m\hbar\omega) f^\eta_{T_d,\mu_d}(\varepsilon_{c\pm}+m\hbar\omega),
\end{equation}
where $f^*_{T,\mu}(E)=f_{T,\mu}(E)$, $f^\dag_{T,\mu}(E)=1-f_{T,\mu}(E)$ 
as well as $\gamma_{cd,\pm}^{(m)}=\frac{1}{T}\int_0^T\!\mathrm{d}t\,e^{im\omega t}
\langle\text{vac}|\gamma\ao_{cd}| u_{c\pm}(t)\rangle$. 
For the undriven left dot only one term of the sum contributes, 
$\gamma_{cl,\pm}^{(m)}=\gamma \delta_{m+s_c\nu}/\sqrt{2}$, describing the 
coupling to reservoir states at energies 
$E_0+s_c (E_g+\Delta)\pm J_c^\text{eff}$. 
In contrast for the driven right dot, we find coupling matrix elements 
$\gamma_{cr,\pm}^{(m)}=\gamma \mathcal{J}_{m}(\alpha)/\sqrt{2}$
for particle exchange with reservoir states at all energies
$E_0+s_c E_g \pm J_c^\text{eff}+m\hbar\omega$. The ``satellite'' coupling terms 
with non-zero $m$ are a direct consequence of the periodic 
time-dependence of the Floquet states $|u_{c\pm}(t)\rangle$ known as 
\emph{micromotion}. 

For small driving amplitudes one has
$\mathcal{J}_{m}(\alpha)\simeq (\alpha/2)^m/m!$. 
Thus, the ideal situation captured by Eqs.~(\ref{eq:Tratio}) and
(\ref{eq:pumpregime}), where the right dot is just coupled to reservoir states 
of energy $E_0+s_c E_g$, is given in the limit $\alpha\to0$, only. However, in
this limit also the effective tunneling matrix elements vanish, 
$J_c^\text{eff}\to 0$, suppressing transport between both reservoirs. As a 
result, when choosing $\alpha$ there will be a trade-off between enhancing 
$J_c^\text{eff}$ and avoiding detrimental processes associated with 
photon-assisted ($m\ne0$) tunneling between the system and the right reservoir.
Since our theory is limited to the regime $\gamma\ll|J_c^\text{eff}|$, where the 
bottleneck of transport from one reservoir to the other through the double dot
is $\gamma$ rather than $J_c^\text{eff}$, it does not directly describe the 
suppression of transport for $\alpha\to0$. However, it is still taken into 
account indirectly by the fact that transport is limited by a value of
$\gamma$, which is assumed to be smaller than $|J_c^\text{eff}|$. 

In the steady state the mean-occupations of the Floquet modes obey
$\frac{\mathrm{d}}{\mathrm{d}t} \langle\no_{c\pm}\rangle
= R^{*}_{c\pm}(1-\langle\no_{c\pm}\rangle) 
- R^{\dag}_{c\pm} \langle\no_{c\pm}\rangle =0$,
so that 
$
\langle\no_{c\pm}\rangle 
										= (1+R^{\dag }_{c\pm}/R^{*}_{c\pm})^{-1} 
$. 
From this solution, we obtain the steady-state rates for the change of particle 
number, $\dot{N}_d$, and energy, $\dot{E}_d$, in both reservoirs:
\begin{eqnarray}\label{eq:NEdot}
\dot{N}_d&=&\sum_{c\pm}\!\left[ R_{c\pm}^{\dag d} \langle\no_{c\pm}\rangle
						-R_{c\pm}^{* d} (1-\langle\no_{c\pm}\rangle) \right]
\\\nonumber
\dot{E}_d&=& \sum_{c\pm\atop m}\!
			\left[ R_{c\pm}^{\dag d(m)} \langle\no_{c\pm}\rangle
						-R_{c\pm}^{* d(m)} (1-\langle\no_{c\pm}\rangle) \right]
		\big(\varepsilon_{c\pm}+m\hbar\omega\big).
\end{eqnarray}
While the total particle number is conserved, so that $\dot{N}_r+\dot{N}_l=0$, 
energy is not conserved for non-zero driving, so that 
$\dot{E}_r+\dot{E}_l\ne 0$ (we always find $\dot{E}_r+\dot{E}_l> 0$, 
consistent with the second law of thermodynamics). In the quantum-heat-pump 
regime, $\dot{E}_r<0$ for $T_r<T_l$, where the system extracts energy from the
colder right reservoir, we can define the coefficient of performance 
$\text{COP}=-\dot{E}_r/(\dot{E}_r+\dot{E}_l)$. 

\section{Reservoir dynamics}

Assuming the reservoirs to thermalize sufficiently fast to individually remain 
in equilibrium while exchanging energy and particles with the double dot, we
compute the time derivatives of $\mu_d$ and $T_d$ from Eqs.~(\ref{eq:NEdot}). 
For this purpose, we invert
\begin{equation}
\begin{pmatrix}
{\dot N}_d\\
{\dot E}_d
\end{pmatrix}
=
\begin{pmatrix}
\partial_{\mu_d} N_d & \partial_{T_d}  N_d\\
\partial_{\mu_d} E_d & \partial_{T_d}  E_d
\end{pmatrix}
\begin{pmatrix}
{\dot \mu_d}\\
{\dot T_d}
\end{pmatrix},
\label{eqmat}
\end{equation}
where $N_d=N_d(T_d,\mu_d) = \int_0^\infty\!\mathrm{d}\epsilon \rho_d
f_{T_d,\mu_d}(\epsilon)$ and $E_d=E_d(T_d,\mu_d) = \int_0^\infty\!\mathrm{d}\epsilon
\rho_d\epsilon f_{T_d,\mu_d}(\epsilon)$. In the following, we will use 
the tunneling parameter $J$ as the unit of energy and measure times in 
units of $\tau = \hbar/(2\pi\gamma^2\rho_r)$, so that $\gamma$ drops out and 
the dynamics depends on the ratio $\lambda \equiv \rho_l/\rho_r$ of reservoir 
``sizes'' rather than on the absolute values $\rho_r$ and $\rho_l$.


\begin{figure}
\begin{center}
  \includegraphics[width=0.45\linewidth]{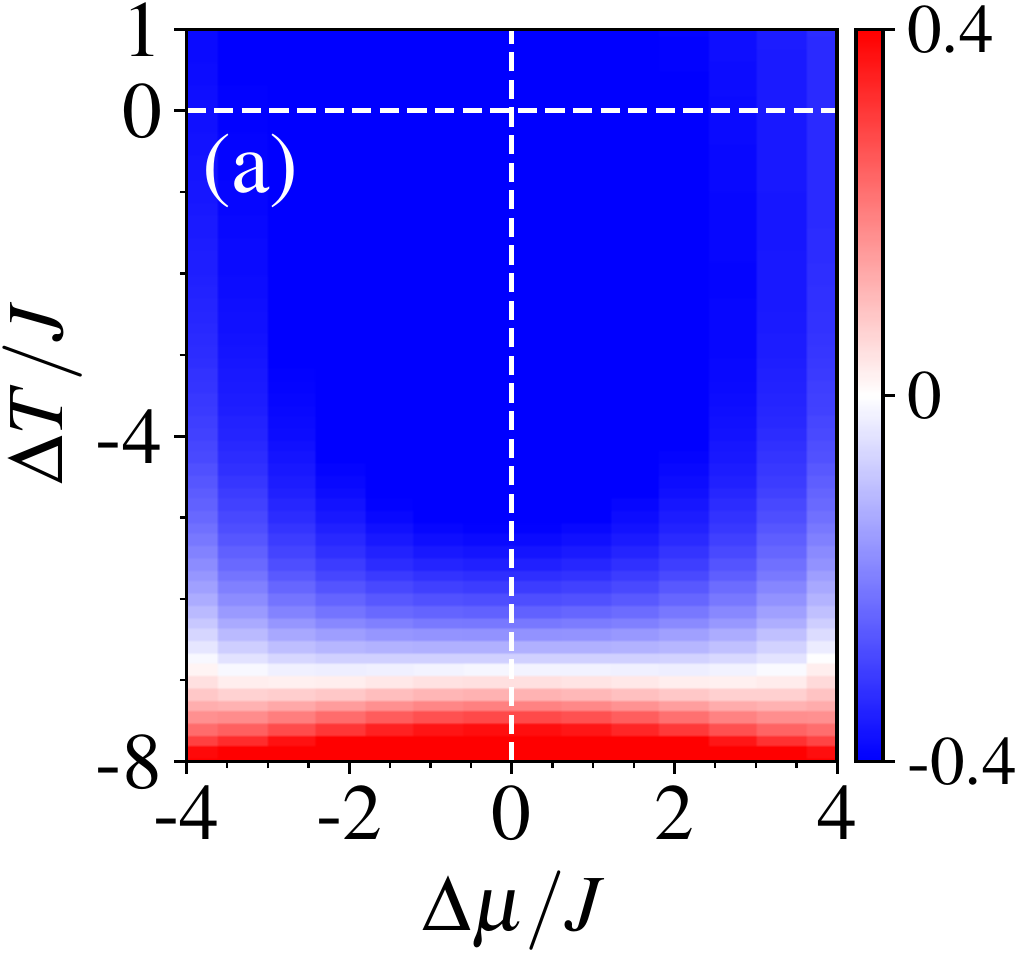}
  \hspace{0.05cm}
  \includegraphics[width=0.52\linewidth]{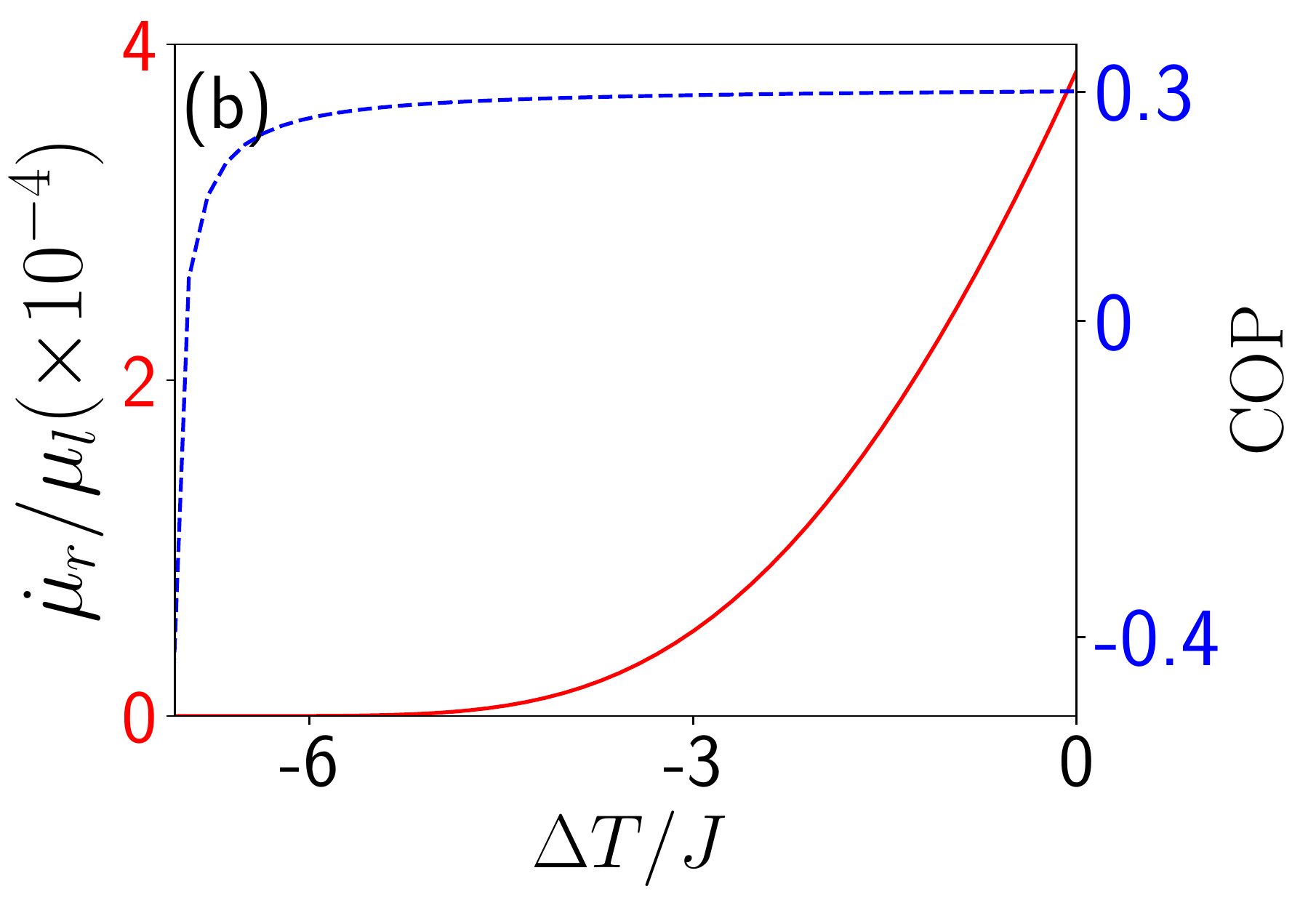}
\end{center}
   \caption{(a) Change rate of the right temperature, ${\dot T}_r$,
vs.\ $\Delta T=T_r-T_l$ and $\Delta\mu=\mu_r-\mu_l$, for fixed $\mu_l=E_0=46$ and $T_l=0.2\mu_l$ and $\alpha = 0.1$, $\Delta=\hbar \omega = 20$, $E_g=6$, 
$\lambda=10$. 
(b) Relative change rate of right chemical potential ${\dot \mu}_r/\mu_r$ (red solid line, left axis) and coefficient of performance $\text{COP}$
vs.~$\Delta T$ (blue dotted line, right axis) for $\Delta\mu=0$ for parameters 
of (a).}
\label{tr}
\end{figure}

\section{Results}
In order to test, whether the driven double dot can operate as a heat pump, in 
Fig.~\ref{tr}(a), we examine the rate $\dot{T}_r$ at which the right 
temperature changes in response to finite differences in temperature and 
chemical potential, $\Delta T=T_r-T_l$ and $\Delta\mu=\mu_r-\mu_l$, keeping 
$T_l$ and $\mu_l$ fixed (the parameters are given in the caption). Since we 
are considering a small 
driving amplitude $\alpha=0.1$, for $\Delta \mu=0$ we expect the system to 
operate as a heat pump roughly in the regime (\ref{eq:pumpregime}), i.e.\ for
$0>\Delta T>-(1-1/a)T_l\approx-7.1$. And, indeed, we find $\dot{T}_r<0$ for
$\Delta T\gtrsim -7$, in excellent agreement with this prediction. Moreover, 
the relative rate at which the right chemical potential changes
[Fig.~\ref{tr}(b), red line] is negligibly small, as desired for the operation 
of the device as a heat pump. The blue line in Fig.~\ref{tr}(b) shows the 
coefficient of performance, reaching its maximum of about $0.3$ at $\Delta T=0$. 

\begin{figure}
\includegraphics[width=0.8\linewidth]{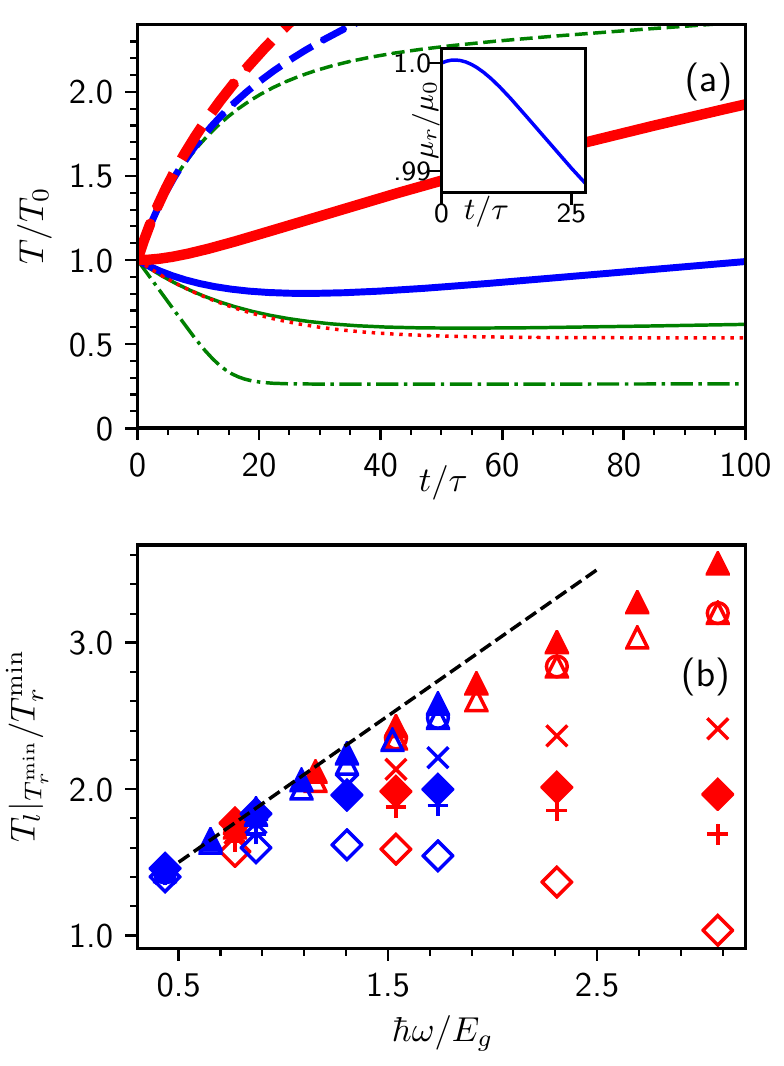}
\caption{(a) Time evolution of $T_r$ (solid lines) and $T_l$ (dashed lines) 
for initial conditions $T_r=T_l=T_0=0.2\mu_l$ and $\mu_l=\mu_r=E_0=46$;
driving amplitudes $\alpha=0.1$ (thin green), $0.25$ (intermediate 
blue), $0.5$ (thick red); as well as for $E_g=6$, $\Delta=\hbar\omega=20$ 
(giving $a=13/3$), and $\lambda=1$. 
Dotted red line shows $T_r$ for $\alpha=0.5$ neglecting rates
$R_{c\pm}^{\eta d(m)}$ with $m\ne0$. Dot-dashed green line shows $T_r$ for
$\alpha=0.1$ and $\lambda=10$.
Inset: Evolution of right chemical potential for $\alpha =0.25$. 
(b) $T_l/T_r$ when $T_l$ reaches its minimum vs.\ $\hbar\omega/E_g$; 
for $E_g=6$ (red symbols), $=12$ (blue symbols); different 
$\hbar\omega=\Delta$; $\alpha=0.1$ (triangles), $0.2$ (circles),
$0.3$ (diagonal crosses), $0.4$ (pluses), $0.5$ (diamonds);   
$\lambda=10$ (filled symbols), $1$ (open and other symbols). The dashed line 
shows $a$.}  
\label{tkchange}        
\end{figure}

In Fig.~\ref{tkchange}(a), we plot the time evolution of $T_r$ (solid lines) 
and $T_l$ (long-dashed lines) starting from the same initial temperature $T_0$ 
and chemical potential $\mu_0=E_0$ (the parameters are given in the caption).
The thin green, intermediate blue, and thick red lines correspond to increasing 
driving strengths $\alpha=0.1$, $0.25$, and $0.5$, respectively. The best 
performance can be observed for the weakest driving strength. Here the 
temperature of the left reservoir first drops to $0.6 T_0$ due to heat 
transfer to the left reservoir, before it increases again very slowly as a 
result of driving induced heating. Increasing $\alpha$ and with that also
$|J_c^\text{eff}|$ the dynamics can be made faster by increasing $\gamma$ so 
that $\tau$ becomes smaller. However, at the same time also driving-induced 
heating increases with $\alpha$ and causes a smaller temperature reduction to
$0.8 T_0$ for $\alpha=0.25$ and even a temperature increase of both 
reservoirs for $\alpha=0.5$. The inset shows the time evolution of the 
chemical potential $\mu_r$ of the right reservoir for $\alpha=0.25$. We can see 
that it changes by about one percent only on the time scale needed to reach the 
minimum temperature, so that the driven double dot predominantly cools the 
right reservoir. 

The driving-induced heating is directly associated with the micromotion and the 
resulting rates $R^{\eta d(m)}_{c\pm}$ with $m\ne0$ [Eq.~(\ref{eq:Rm})]. When 
we set the $m\ne0$-rates to zero artificially by hand, no driving-induced 
heating occurs and a steady state with $T_r\approx 0.54T_0$ is reached, as can 
be seen from the thin short-dashed red line showing $T_r$ for $\alpha=0.5$
(where before no cooling was observed at all). For this artificial steady 
state, we expect $T_l/T_r\approx 1+\Delta/E_g=a$, according to
Eq.~(\ref{eq:Tratio}). Including also $m\ne0$, this estimate sets also an upper 
limit for $T_l/T_r$ at the time where $T_r$ becomes minimal. This is confirmed 
in Fig.~\ref{tkchange}(b), where we plot this ratio versus 
$\Delta/E_g=\hbar\omega/E_g$ for various different parameters (different 
symbols, see caption) together with $a$ (dashed line). From this Figure, we can 
see that, as expected, the data comes closer to the optimal limit $a$ (dashed 
line) when $\alpha$ is lowered. 
While also lowering $\hbar\omega/E_g$ helps to approach the bound $a$, the bound
itself becomes more favorable for larger $\hbar\omega/E_g$, so that (at least
for sufficiently small $\alpha$) we still find better (larger) temperature 
ratios for larger $\hbar\omega/E_g$.

\begin{figure}
 \includegraphics[width=8.5cm]{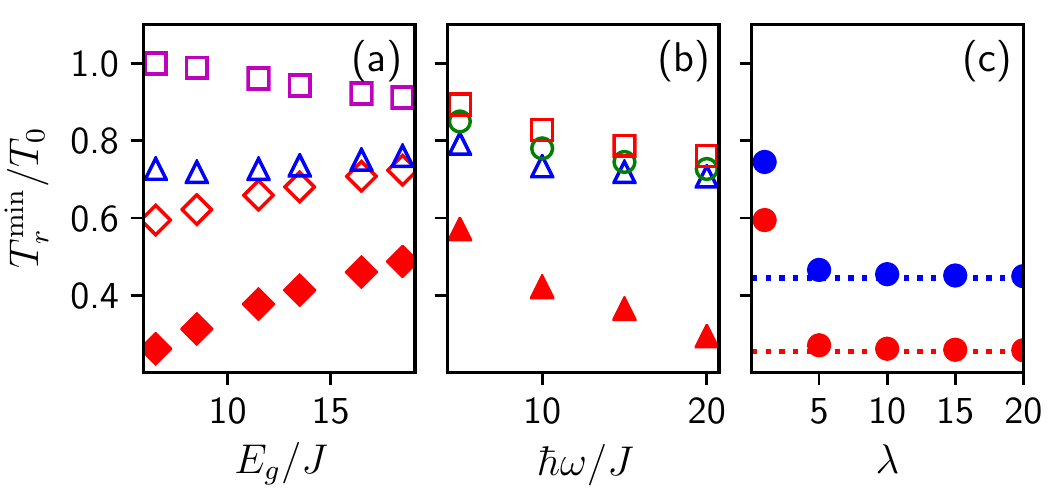}
  \caption{$T_r^\text{min}/T_0$ for 
	initial conditions $T_r=T_l=T_0=0.2\mu_l$ and $\mu_l=\mu_r=E_0=46$ 
	(a) vs.\ $E_g$ for $\hbar\omega/J=20$; $\alpha = 0.1$ (diamonds), 
	$0.2$ (triangles), $0.5$ (squares); $\lambda=1$ (open symbols), $10$ (filled symbols);
	(b) vs.\ $\hbar\omega=\Delta$ for $\alpha=0.2$; $E_g = 6$ (triangles), 
	$12$ (circles), $18$ (squares); $\lambda=1$ (open symbols), $10$ (filled symbols);
	(c) vs.\ $\lambda$ for $E_g=6$ for $\omega=\Delta=20$, $\alpha=0.1$ (lower red circles) and $E_g=12$, $\omega=\Delta=15$, $\alpha=0.2$ (upper blue
	circles); the dashed lines show the corresponding results for
  $\lambda\to\infty$.}
   \label{trmin}        
\end{figure}

Finally, we observe larger temperature ratios when increasing the relative size 
$\lambda$ of the energy absorbing left bath. This is a consequence of the fact 
that the left bath can absorb more of the heating-induced energy during the 
evolution, since its temperature remains lower. An additional major benefit of 
raising $\lambda$ is, furthermore, that not only $T_r/T_l$ is lowered, but at 
the same time also the absolute values of $T_l$ and with this also $T_r$. The 
advantage of an increased left reservoir is confirmed by the dash-dotted green 
line in Fig.~\ref{tkchange}(a) showing the evolution of $T_r$ for the same 
parameters like the solid green curve, except that now $\lambda=10$ is chosen 
rather than the value $\lambda=1$ used for all other curves. As a result, $T_r$ 
is reduced to $0.26 T_0$, which is almost the optimal value 
$T_0/a \approx 0.23 T_0$ of the bound 
\begin{equation}\label{eq:bound}
T_r\gtrsim T_0/a
\end{equation}
obtained from Eq.~(\ref{eq:pumpregime}) for $\lambda\to\infty$ so that
$T_l=T_0$ at all times. 
In Fig.~\ref{trmin} we investigate how the minimal right temperature assumed 
during the evolution depends on the system parameters. As expected from 
Eq.~(\ref{eq:Tratio}), we find that $T_r^\text{min}/T_0$ decreases both when 
lowering $E_g$ [at least for sufficiently weak driving strength $\alpha$, 
Fig.~\ref{trmin}(a)] and when increasing $\hbar\omega=\Delta$ 
[Fig.~\ref{trmin}(b)]. Moreover, we find that the enhancement of cooling by 
increasing the relative size $\lambda$ of the right reservoir saturates at 
values of about $\lambda=5$ [Fig.~\ref{trmin}(c)]. This is a very promising 
result for the implementing such a heat pumps with ultracold atoms, since it 
implies that it is sufficient to engineer reservoirs of rather moderate size.

\begin{figure}
    \centering
    \includegraphics[width=1\linewidth]{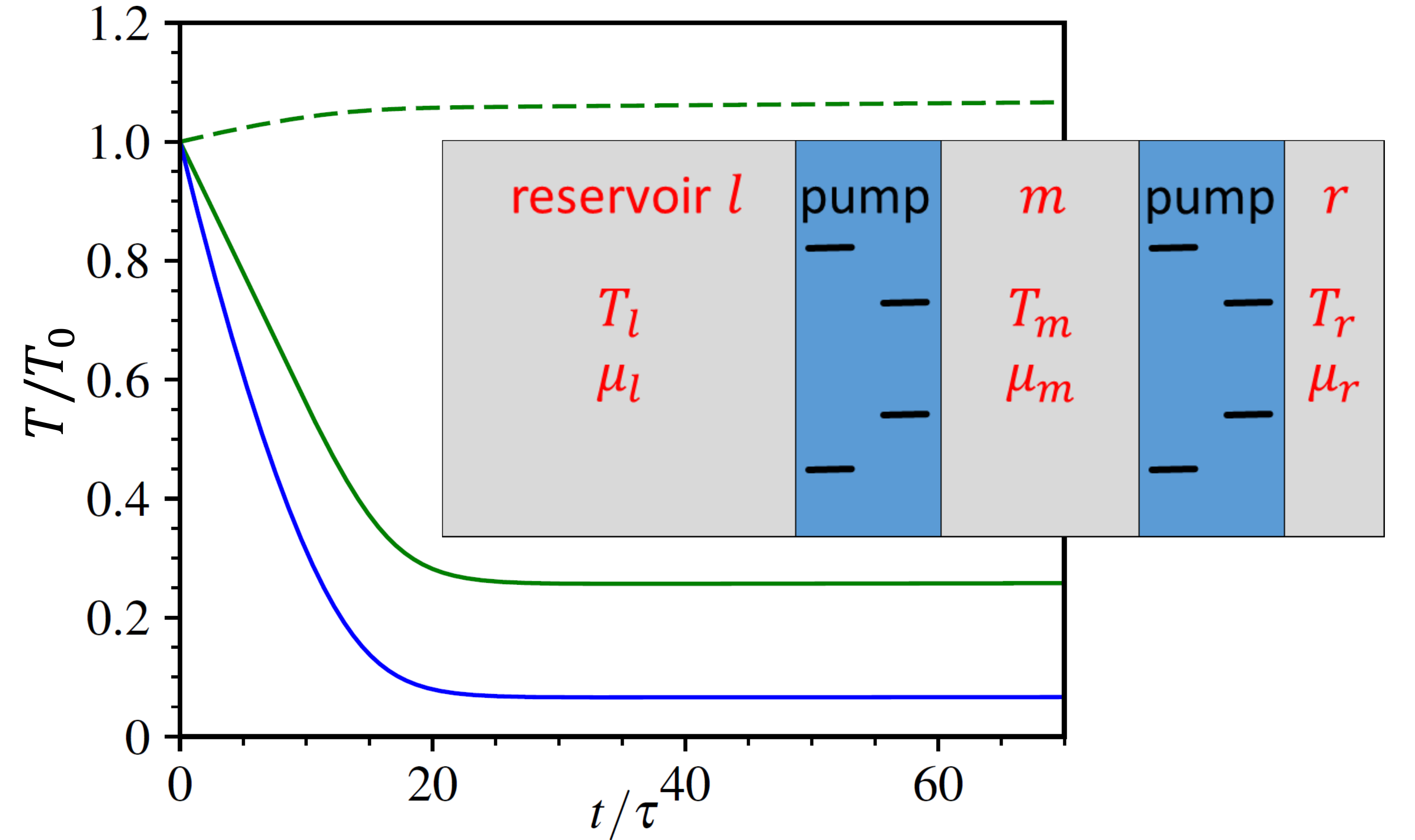}
    \caption{Temperature evolution of three reservoirs $l$, $m$, $r$ (upper, middle, lower line) coupled by identical heat pumps; for initial conditions $T_l=T_m=T_r=T_0=0.2\mu_r$, $\mu_l=\mu_m=\mu_r=E_0=46$ and 
		$\hbar\omega=\Delta=20$, $E_g=6$, $\alpha=0.1$, $\rho_l/\rho_m=\rho_m/\rho_r=5$.}
   \label{3rs}
  \end{figure}

Besides the optimization of parameters with the aims of both saturating and 
lowering the bound (\ref{eq:bound}), an alternative strategy for reaching lower 
temperatures is to put two (or more) heat pumps in series. This scenario is 
investigated in Fig.~\ref{3rs}, where we plot the evolution of the temperatures 
of three reservoirs ($l$, $m$, $r$) that are coupled by two identical heat 
pumps (see sketch). Motivated by the results of Fig.~\ref{trmin}(c), we have 
chosen a hierarchy of reservoir sizes according to 
$\rho_l/\rho_m=\rho_m/\rho_r=5$. Note that both left and middle reservoir 
together are still only 30 times larger than the reservoir that we wish to cool.
The lower limit for the temperature of the right reservoir is now given by
$T_r\gtrsim T_0/a^2\approx 0.053 T_0$ and, indeed, we can see that during the 
evolution $T_r/T_0$ is reduced to the temperature $0.07 T_0$, which is only slightly larger. 

\section{Conclusion}
We have described a simple design for a quantum heat pump. It is based on two 
coherently coupled, periodically driven quantum dots, each possessing two 
relevant quantum states and being tunnel-coupled to a reservoir. The working 
principle is based on energy-selective ``photon''-assisted tunneling processes,
where a particle that tunnels from one dot to the other either absorbs or emits 
the energy quantum $\hbar\omega$ associated with the driving frequency, depending
on its energy. We simulate the device using an open-system approach based on 
Floquet-Born-Markov theory and show that it indeed works as a heat pump and that
unavoidable fundamental driving-induced heating is directly linked to the
micromotion of the system's Floquet states. It is a promising perspective to
implement such a device with quantum-gases using recently established 
experimental techniques for microstructuring light-shift potentials. Namely, 
quantum gases provide extremely clean conditions, since they do not suffer from
detrimental dissipation via radiative losses or the coupling to a phonon-bath, 
as it is present in electronic systems. Moreover, the proposed device might also
be of practical use for reaching lower temperatures.


\begin{acknowledgments}
We acknowledge insightful discussions with Manuel Alamo, Martin Lebrat, Laura 
Corman, and Daniel Vorberg. This research was funded by the Deutsche 
Forschungsgemeinschaft (DFG) via the Research Unit FOR 2414 under Project
No.\ 277974659. \end{acknowledgments}

\bibliography{hpump}{}
\bibliographystyle{apsrev4-1}
\end{document}